\documentclass[aps,prd,onecolumn,preprintnumbers,groupedaddress,showpacs,nofootinbib,amssymb]{revtex4-2}
\usepackage{graphicx,color}
\usepackage{amsmath}
\usepackage{amssymb}
\usepackage{amsfonts}
\usepackage{bm}
\usepackage{cancel}
\usepackage{slashed}
\newcommand{\e}{\mathrm{e}}

\allowdisplaybreaks[4]
\begin{document}

\preprint{KEK-TH-2802, KEK-Cosmo-0409}
\title{May Negative Mass Objects exist in the sky?}
\author{Shin'ichi~Nojiri,$^{1,2}$}
\email{nojiri@nagoya-u.jp}
\author{S.D. Odintsov,$^{3,4}$}
\email{odintsov@ieec.cat}
\affiliation{$^{1)}$ Theory Center, High Energy Accelerator Research Organization (KEK), \\
Oho 1-1, Tsukuba, Ibaraki 305-0801, Japan \\
$^{2)}$ Kobayashi-Maskawa Institute for the Origin of Particles
and the Universe, Nagoya University, Nagoya 464-8602, Japan \\
$^{3)}$ ICREA, Passeig Luis Companys, 23, 08010 Barcelona, Spain\\
$^{4)}$ Institute of Space Sciences (IEEC-CSIC) C. Can Magrans s/n, 08193 Barcelona, Spain\\
}

\begin{abstract}
We conjecture the possibility of negative mass objects (NMOs) existing in the sky. 
It is shown that they may not be so exotic as usually expected.
We show that NMOs appear as solutions of standard gravitational equations if we consider the system of a compact positive mass object, cosmological fluid and negative cosmological constant. 
We also construct models which generate such NMOs as solutions within the two-scalar theory and scalar-Einstein-Gauss-Bonnet gravity inspired by string theory. 
The orbits of the photon and massive particles are investigated in the background, where there is a negative mass object which realises a kind of effective anti-gravity.
It is explicitly found that the bound system consisting of a positive mass object and a negative mass object can be formed in spite that a positive mass object suffers the repulsive force from the NMO. 
The possibility that such exotic objects might be observed is discussed. 
A simple conjecture about their possible masses is made, too.
As an even more exotic object, we consider a non-trivial object with vanishing mass and investigate its properties. 
\end{abstract}

\maketitle

\section{Introduction}\label{intro}

Since the mathematical formulation of gravity, the possibility of anti-gravity attracted a lot of attention in science fiction literature as well as in scientific literature. 
Different attempts to realise anti-gravity theoretically were done recently, see, for instance,~\cite{Oikonomou:2014sra, Oikonomou:2014dga, Oltean:2014bua, Bamba:2014kza, Carrasco:2013hua, Bars:2012fq, Dolgov:2012qf, Bars:2011aa} and Refs. therein. 
If we assume that Newton's gravity is valid, when there are two point masses with the position vectors $\bm{r}_1$ and $\bm{r}_2$ and the masses $m_1$ and $m_2$, the force that a point mass with mass $m_1$ suffers from another point mass with mass $m_2$ is given by 
\begin{align}
\label{eomForce}
\bm{F}_{2\to 1} = - \frac{G m_1 m_2 \left( \bm{r}_1 - \bm{r}_2 \right)}{\left| \bm{r}_1 - \bm{r}_2 \right|^3}\, . 
\end{align}
The anti-gravity means that $G$ is negative, but if $G$ is negative, then the Einstein-Hilbert action is given by 
\begin{align}
\label{EH}
R_\mathrm{EH}=\frac{1}{2\kappa^2} \int d^4 x \sqrt{-g} R \, , \quad 
\kappa^2 = 8\pi G\, ,
\end{align}
the graviton becomes a ghost, and the theory becomes physically inconsistent. 
We should note that the appearance of the ghost does not mean the instability of the system, but it tells us that the theory is physically inconsistent. 
Of course, one can conjecture on the effective change of gravitational coupling sign or just effective anti-gravity regions in the very early universe, as was done in some of the above papers. 
Nevertheless, it is hard to consistently realise such a phenomenon even on a theoretical level.

In this work, we try to develop another approach based on the effective anti-gravity, by which we mean the gravitational force produced by NMOs. 
Starting from the case where the negative mass means $m_1$ and/or $m_2$, we show that it does not always lead to any inconsistency. 
Note that there are several attempts to consider NMOs, see~\cite{Kreinovich:2018old, Klinkhamer:2018dta, Okunev:2024cso, Manfredi:2026qoz} and Refs.~therein. 
We now also consider the possibility that there might exist objects with negative mass. 

A natural question could be whether such an object could exist consistently. 
The Lagrangian of a particle at $\bm{r}$ with mass $m$ and the potential $V\left(\bm{r}\right)$ in classical mechanics is given by 
\begin{align}
\label{clsclprtcl}
L=\frac{1}{2} m \bm{r}\cdot \bm{r} - V\left(\bm{r}\right) \, .
\end{align}
If mass $m$ is negative, a ghost appears in the corresponding quantum mechanics. 
We may consider, however, how a bubble inside water behaves. 
The bubble behaves as if it has a negative mass because the water around the bubble falls due to gravity. 
There is, of course, no ghost in the system of the bubble and water. 
It could not be so unnatural if we consider NMOs coupled with gravity, because the negative cosmological constant can be regarded as a cosmological fluid with negative energy density. 
The anti-de Sitter spacetime can be realised by the negative cosmological constant, but the expansion of the universe cannot be realised only by the negative cosmological constant. 
If there is another cosmological fluid besides the negative cosmological constant, and the total energy density is positive, the expansion of the universe occurs (for observational indications in such case, see~\cite{Visinelli:2019qqu}). 

We may consider the first FLRW equation, 
\begin{align}
\label{1FLRWeq}
\frac{3}{\kappa^2} H^2 = \rho\, ,
\end{align}
Here $\rho$ is the energy density of a cosmological fluid filling the universe.  
If $\rho$ is negative in a local region, the region can be regarded as having a negative mass. 
The cosmological constant $\Lambda$ can also be regarded as a cosmological fluid with $\rho=\frac{\Lambda}{2\kappa^2}$. 
As is well-known in the anti-de Sitter space, the cosmological constant can be negative. 
If there is another cosmological fluid, whose energy density is denoted by $\rho_f$,  besides the negative cosmological constant, the total energy density is almost equal to zero, 
$\rho_f + \frac{\Lambda}{2\kappa^2} \sim 0$, and therefore the spacetime is almost static. 
The energy density $\rho_f$ can depend on coordinates, $\rho_f=\rho_f\left(x^\mu\right)$ in general. 
If there could be a fluctuation in the fluid and if the density of the fluid becomes smaller in some regions, $\rho_f + \frac{\Lambda}{2\kappa^2} < 0$ due to the fluctuation, the regions behave as if there are NMOs, like the bubble in the water. 
Such fluctuations could not be stable in general. 
We may consider the situation where there is a usual positive point mass, like a compact star, in the fluid with negative pressure, which may appear in the cosmological fluid, like dark energy, and with a negative cosmological constant. 
Due to the negative pressure, the force generated by the gradient of the pressure balances with the gravity generated by the positive point mass, that is, the point mass pushes away the fluid. 
As a result, regions with negative energy density appear. 
If the total mass of the fluid pushed away by the point mass is larger than the mass of the point mass, an object with effectively negative mass is created. 
Such a configuration is stable in general. 

If we linearise the Einstein equation to obtain a Newton approximation, it is easy to find that the NMO gives a repulsive force to the object with a positive mass, that is, the mass is defined by the Newton law of gravity. 
The equivalence principle by Einstein tells us that the NMO also has a negative inertial mass. 
Therefore, the kinetic energy of the NMO is negative, and the kinetic energy becomes increasingly larger if the speed of the NMO becomes larger and larger. 
Of course, if the speed becomes too large, the particle description of the NMO could not be valid. 
Therefore, as long as the description of special relativity is valid, the NMO behaves as a particle with a negative kinetic energy. 

In Section~\ref{eNMO}, we clarify how effectively a negative mass object appears from the system of a positive point mass, cosmological fluid with negative pressure and negative cosmological constant. 
In Section~\ref{CnstrctnMG}, the models realising NMOs in the framework of the two-scalar model and the scalar-Einstein-Gauss-Bonnet gravity inspired by strings are constructed.
In Section~\ref{Phtn}, we investigate the orbit of a photon and show that the photon suffers a repulsive force from the NMO. 
Section~\ref{mssv} is devoted to studying the behaviour of a massive particle and its geodesic around NMO by using the Newton limit. 
In Section~\ref{udrgrdtmch}, by using Newton's gravity and Newton's laws of motion, we investigate the motion of the particles in the presence of NMOs. 
Section~\ref{obsrvtn} is devoted to the development of some conjectures on the observational possibility of finding NMOs. 
In Section~\ref{Svmss}, as another exotic object, we consider and study the properties of the object with vanishing mass.
Some outlook is given in the last Section~\ref{SD}. 

\section{Effectively negative mass object}\label{eNMO}

In this section, we consider the possibility that an NMO might appear when there is a positive massive object in addition to the cosmological fluid with negative pressure and negative cosmological constant. 

\subsection{Behaviour of perfect fluid with negative pressure under the gravitational background}

Let $w$ denote the equation of state (EoS) parameter. 
Note that the compact massive object blows the fluid with $0>w>-\frac{1}{3}$ due to the negative pressure of the fluid. 
To see this, we consider a perfect fluid with a constant EoS parameter $w$ in the Schwarzschild-type static spacetime, 
\begin{align}
\label{GBiv}
ds^2 = \sum_{\mu,\nu=t,r,\theta,\varphi} g_{\mu\nu} dx^\mu dx^\nu =&\, - \e^{2\nu (r)} dt^2 + \e^{2\lambda (r)} dr^2 
+ r^2 \sum_{i,j=\theta,\varphi} {\bar g}_{ij} dx^i dx^j \, , \nonumber \\
\sum_{i,j=\theta,\varphi} {\bar g}_{ij} dx^i dx^j =&\, d\theta^2 + \sin^2\theta \, d\varphi^2 \, .
\end{align}
Let $\rho$ and $p$ be the total energy density and the pressure of the fluid, respectively, and satisfy the following conservation law, 
\begin{align}
\label{FRN2}
0 = \nabla^\mu T_{\mu r} = \nu' \left( \rho + p \right) + \frac{dp}{dr} \, .
\end{align}
Here $\rho$ and $p$ only depend on the radial coordinate $r$. 
By using the EoS $p=w\rho$, Eq.~\eqref{FRN2} can be integrated to give, 
\begin{align}
\label{NM1}
\rho=\rho_0 \e^{- \frac{1+w}{w}\nu}\, .
\end{align}
Here $\rho_0$ is a constant of the integration. 
In the case of the Schwarzschild metric, 
\begin{align}
\label{Schwrzschld}
\e^{2\nu}=\e^{-2\lambda}=1 - \frac{2M}{r}\, , 
\end{align}
$\nu$ is a monotonically increasing function of $r$. 
If $0>w>-1$, $\rho$ is also an increasing function of $r$, that is, if $r$ becomes smaller, the density becomes smaller, that is, the fluid is displaced by the compact massive object. 

We now consider the massive compact object whose radius is $2M$.
The lost mass $M_\mathrm{lost}$ could be given by 
\begin{align}
\label{lmss}
M_\mathrm{lost} = 4\pi \int_{2M}^\infty dr r^2 \left( \rho (r=\infty) - \rho(r) \right) 
= 4\pi \rho_0 \int_{2M}^\infty dr r^2 \left\{ 1 - \left( 1 - \frac{2M}{r} \right)^{-\frac{1+w}{2w}} \right\} \, ,
\end{align}
which diverges when $r\to \infty$ and we need a cutoff for large $r$, $r=r_\Lambda$. 
Such a cutoff could be natural due to the structure of the universe, but we need to consider what could be a natural cutoff scale. 
For example, the number density of the massive compact object is $n$, we find $r_\Lambda\sim n^{-\frac{1}{3}}$. 
Anyway, we find $M_\mathrm{lost}\gg M$ and therefore the lost mass is much larger than the mass of the massive compact object, and there appears an effectively negative mass object. 

By changing the variable $r$ to $s$ by 
\begin{align}
\label{rs}
s=1 - \frac{2M}{r}\, , 
\end{align}
we find 
\begin{align}
\label{rs2}
ds = \frac{2M}{r^2} dr\, , \quad 
r = \frac{2M}{1-s} \, , \quad dr = \frac{2M}{\left(1-s\right)^2}ds \, ,
\end{align}
and 
\begin{align}
\label{lmss2}
M_\mathrm{lost} 
= 4\pi \rho_0 \int_0^{1 - \frac{2M}{r_\Lambda}} ds \frac{8M^3}{\left(1-s\right)^4} \left\{ 1 - s^{-\frac{1+w}{2w}} \right\} \, .
\end{align}
Especially when $w=-\frac{1}{3}$, that is, $-\frac{1+w}{2w}=1$, we find 
\begin{align}
\label{lmss3}
M_\mathrm{lost} 
= 16\pi \rho_0 M^3 \left\{ \frac{{r_\Lambda}^2}{4M^2} - 1 \right\} \, .
\end{align}
This could be an explicit example of a stable NMO. 

We may obtain a finite result for the lost mass $M_\mathrm{lost}$ without using the cutoff scale $r=r_\Lambda$ but by modifying the EoS. 
Although rather artificial, one may assume the following example of the EoS, 
\begin{align}
\label{nlEoS}
p = - \rho_0 \left\{ \frac{4}{3} - \frac{4}{3} \left( 1 - \frac{\rho}{\rho_0}\right)^\frac{1}{4} - \frac{6}{5} \left( 1 - \left( 1 - \frac{\rho}{\rho_0}\right)^\frac{1}{4} \right)^2 
+ \frac{4}{7} \left( 1 - \left( 1 - \frac{\rho}{\rho_0}\right)^\frac{1}{4} \right)^3 
 - \frac{1}{9} \left( 1 - \left( 1 - \frac{\rho}{\rho_0}\right)^\frac{1}{4} \right)^4 \right\} \, .
\end{align}
With the the conservation law~\eqref{FRN2}, one finds 
\begin{align}
\label{nlrho}
\rho= \rho_0 \left( 1 - \left( 1 - \e^{2\nu} \right)^4 \right) \, .
\end{align}
By using the expression \eqref{nlrho} with the Schwarzschild metric in \eqref{Schwrzschld}, instead of \eqref{lmss}, we find the following expression of the lost mass $M_\mathrm{lost}$, 
\begin{align}
\label{lmssnl}
M_\mathrm{lost} = 4\pi \int_{2M}^\infty dr r^2 \left( \rho (r=\infty) - \rho(r) \right) 
= 4\pi \rho_0 \int_{2M}^\infty dr r^2 \left( \frac{2M}{r} \right)^4 
= 32\pi M^3\rho_0 \, . 
\end{align}
Then if $M_\mathrm{lost}>M$, that is, 
\begin{align}
\label{nlttlmss}
32\pi M^2\rho_0 > 1 \, ,
\end{align}
the total mass becomes negative, 
\begin{align}
\label{tmss}
M_\mathrm{total} = M - M_\mathrm{lost} = M - 32\pi M^3\rho_0\, .
\end{align}
This could be another example of a stable NMO.

\subsection{Negative mass object from perfect fluid}

It is assumed above the Schwarzschild background~\eqref{Schwrzschld} is given, but in a more realistic scenario, we need to consider the back reaction from the fluid to the geometry, that is, we need to solve the full Einstein equation with the conservation law~\eqref{FRN2}. 

For the action of Einstein's gravity with a negative cosmological constant $\Lambda<0$, which is given by 
\begin{align}
\label{JGRG6}
S_\mathrm{EH}=\int d^4 x \sqrt{-g} \left(
\frac{R}{2\kappa^2} - \frac{\Lambda}{2\kappa^2} + \mathcal{L}_\mathrm{matter} \right)\, ,
\end{align}
by assuming the metric in the form of \eqref{GBiv}, the Einstein equations give, 
\begin{align}
\label{FRN4}
 - \kappa^2 \rho =&\, - \frac{1}{2} \left( R - \Lambda \right) - \e^{- 2 \lambda} \left\{
\nu'' + \left(\nu' - \lambda'\right)\nu' + \frac{2\nu'}{r}\right\} \, ,\\
\label{FRN5}
 - \kappa^2 p =&\, \frac{1}{2} \left( R - \Lambda \right) + \e^{ -2\lambda} \left\{ \nu'' + \left(\nu' - \lambda'\right)\nu' 
 - \frac{2 \lambda'}{r} \right\} \, ,\\
\label{FRN6}
 - \kappa^2 p =&\, \frac{1}{2} \left( R - \Lambda \right) - \frac{1}{r^2} \left[ 1 + \left\{ - 1 - r \left(\nu' 
 - \lambda' \right)\right\} \e^{-2\lambda}\right] \, . 
\end{align}
Here, the scalar curvature $R$ is given by 
\begin{align}
\label{R}
R= \e^{-2\lambda}\left\{ - 2\nu'' - 2\left(\nu' - \lambda'\right)\nu' - \frac{4\left(\nu' - \lambda'\right)}{r} + \frac{2\e^{2\lambda} - 2}{r^2} \right\} \, .
\end{align}
Note that Eqs.~\eqref{FRN4}, \eqref{FRN5}, and \eqref{FRN6} reproduce the conservation law in Eq.~\eqref{FRN2}. 
Then one may forget \eqref{FRN6}. 

By substituting the expression of $R$ in \eqref{R} into \eqref{FRN4} and \eqref{FRN5}, we obtain, 
\begin{align}
\label{FRN4B}
 - \kappa^2 \left( \rho + \frac{\Lambda}{2\kappa^2} \right) 
=&\, - \e^{-2\lambda}\left( \frac{2\lambda'}{r} + \frac{\e^{2\lambda} - 1}{r^2} \right) \, ,\\
\label{FRN5B}
 - \kappa^2 \left( p - \frac{\Lambda}{2\kappa^2} \right)
 =&\, \e^{-2\lambda}\left( - \frac{2 \nu'}{r} + \frac{\e^{2\lambda} - 1}{r^2} \right) \, . 
 \end{align}
The Schwarzschild mass is defined by 
\begin{align}
\label{Smass}
m(r) \equiv \frac{4\pi r}{\kappa^2} \left( 1-\e^{-2\lambda(r)} \right) \, .
\end{align}
Because 
\begin{align}
\label{eqs}
m'(r) = \frac{8\pi}{\kappa^2} \left[ \frac{1}{2} \left( 1-\e^{-2\lambda} \right) + r\lambda' \e^{-2\lambda} \right] \, , \quad 
\e^{-2\lambda(r)} = 1 - \frac{\kappa^2 m(r)}{4\pi r} \, ,
\end{align}
by deleting $\lambda$ in \eqref{FRN4B}, we obtain the first Tolman-Oppenheimer-Volkoff (TOV) equation, 
\begin{align}
\label{TOV1}
m'(r)= 4 \pi r^{2} \left( \rho + \frac{\Lambda}{2\kappa^2} \right) \, .
\end{align}
Eq.~\eqref{FRN5B} can be rewritten as 
\begin{align}
\label{nuprime}
\nu'=\frac{r}{2\left( 1 - \frac{\kappa^2 m(r)}{4\pi r}\right)} \left\{ \frac{\kappa^2 m(r)}{4\pi r^3 } + \kappa^2 \left( p - \frac{\Lambda}{2\kappa^2} \right) \right\} \, .
\end{align}
By using \eqref{nuprime}, we delete $\nu'$ in \eqref{FRN2} and obtain, 
\begin{align}
\label{convp}
0 = \frac{r\left( \rho + p \right)}{2\left( 1 - \frac{\kappa^2 m(r)}{4\pi r}\right)} \left\{ \frac{\kappa^2 m(r)}{4\pi r^3 } + \kappa^2 \left( p - \frac{\Lambda}{2\kappa^2} \right) \right\} + \frac{dp}{dr} \, , 
\end{align}
and by using the EoS $p=w\rho$, we find the equation corresponding to the second TOV equation, 
\begin{align}
\label{TOV2}
0 = \frac{r\left(1 + w \right)\rho}{2w \left( 1 - \frac{\kappa^2 m(r)}{4\pi r}\right)} 
\left\{ \frac{\kappa^2 m(r)}{4\pi r^3 } + \kappa^2 \left( w \rho - \frac{\Lambda}{2\kappa^2} \right) \right\} + \rho' \, .
\end{align}
One can find the profile of the spherically symmetric object by solving Eqs.~\eqref{TOV1} and \eqref{TOV2} by regarding $m$ and $\rho$ as independent variables. 

We first solve Eqs.~\eqref{TOV1} and \eqref{TOV2} when $r$ is large by assuming $\rho=\rho_0 r^\alpha$ with constants $\rho_0$ and $\alpha$. 
\begin{itemize}
\item If we assume $\alpha>0$, Eq.~\eqref{TOV1} gives 
\begin{align}
\label{mas}
m \sim \frac{4\pi \rho_0}{3+\alpha} r^{3 + \alpha} \, ,
\end{align}
and Eq.~\eqref{TOV2} also gives, 
\begin{align}
\label{TOV2B}
0= - \frac{\left(1 + w \right)^2 \left(3 + \alpha \right)}{2w } + \alpha \, ,
\end{align}
which can be solved as 
\begin{align}
\label{alph}
\alpha = - \frac{3\left( 1 + w \right)^2}{1 + w^2}\, ,
\end{align}
which contradicts with the assumptions $\alpha>0$. 
\item If we assume $0>\alpha$, Eq.~\eqref{TOV1} leads 
\begin{align}
\label{masB}
m \sim \frac{2\pi \Lambda}{3\kappa^2} r^3 \, ,
\end{align}
by using Eq.~\eqref{TOV2}, we find, 
\begin{align}
\label{TOV2C}
0 = - \frac{6 \left( 1 + w \right) \left( \frac{1}{6} - \frac{1}{2} \right) \Lambda\rho_0}{2 w \Lambda}r^{\alpha -1} + \alpha \rho_0 r^{\alpha - 1}\, , \quad
0 = \frac{1 + w}{w} r^{\alpha -1} + \alpha r^{\alpha - 1}\, ,
\end{align}
which requires 
\begin{align}
\label{alph2}
\alpha = - \frac{1+w}{w}\, ,
\end{align}
which contradicts the requirement $0>\alpha$ because we assume $0>w>-1$. 
\item If $\alpha=0$ and $\rho_0 + \frac{\Lambda}{2\kappa^2}=0$, if we write $\rho\sim \rho_0 + \frac{\rho_1}{r}=-\frac{\Lambda}{2\kappa^2}+ \frac{\rho_1}{r}$, 
Eq.~\eqref{TOV1} tells 
\begin{align}
\label{mm}
m \sim 2\pi \rho_1 r^2 \, ,
\end{align}
and by using Eq.~\eqref{TOV2}, we find 
\begin{align}
\label{rh2}
0=\frac{\left( 1 + w \right)^2 \rho_0}{w \kappa^2 \rho_0} \frac{\Lambda}{2\kappa^2}\, ,
\end{align}
which cannot be satisfied. 

If $\rho_0 + \frac{\Lambda}{2\kappa^2}\neq 0$, Eq.~\eqref{TOV1} tells 
\begin{align}
\label{mmm}
m \sim \frac{4\pi}{3} \left( \rho_0 + \frac{\Lambda}{2\kappa^2} \right) r^3 \, .
\end{align}
Therefore Eq.~\eqref{TOV2} leads to 
\begin{align}
\label{rh3}
0= - \frac{3\left( 1 + w \right) \rho_0}{2w \left( \rho_0 + \frac{\Lambda}{2\kappa^2} \right) } 
\left\{ \frac{\kappa^2}{3} \left( \rho_0 + \frac{\Lambda}{2\kappa^2} \right) + \kappa^2 \left( w \rho_0 - \frac{\Lambda}{2\kappa^2} \right) \right\}
= - \frac{3\left( 1 + w \right) \rho_0}{2w \left( \rho_0 + \frac{\Lambda}{2\kappa^2} \right) } 
\left\{ \left( 1 + 3w \right) \rho_0 - \frac{\Lambda}{\kappa^2} \right\} \, .
\end{align}
This tells 
\begin{align}
\label{rh4}
\rho_0 = \frac{\Lambda}{\kappa^2\left( 1 + 3w \right) } \, .
\end{align}
We should note $\rho_0>0$ but $\Lambda<0$ and therefore if $w<-\frac{1}{3}$, there is a consistent solution. 
By combining \eqref{mmm} and \eqref{rh4}, we find 
\begin{align}
\label{mmmB}
m \sim \frac{4\pi}{3} \left( \frac{2 + 3w}{1+3w} \right) \frac{\Lambda}{2\kappa^2} r^3 \, .
\end{align}
Therefore, as long as $w<-\frac{2}{3}$, $m$ becomes negative and an NMO could appear, although the spacetime is not asymptotically flat. 
\end{itemize}
Then one can find the asymptotic solution, and an NMO may appear. 

If there is a system of the negative cosmological constant, a perfect fluid with $0>w>-1$ and usual matter, NMOs can be easily created. 
First, the usual matter makes a compact object like a star. 
Then, due to the negative pressure, an NMO could be created by the gravity of the compact object made of the usual matter. 

\section{Construction by modified gravities}\label{CnstrctnMG}

As the perfect fluid in the last section only gives the asymptotically non-flat spacetime, one may consider modified gravity as an alternative. 
In the scalar-tensor theory, where a scalar $\phi$ minimally couples with gravity and the action is given by 
\begin{align}
\label{scalartensor}
S_\mathrm{scalar-tensor} = \int d^4 x \sqrt{-g} &\, \left\{ \frac{R}{2\kappa^2} - \frac{1}{2} \partial_\mu \phi \partial^\mu \phi - V(\phi) \right\} \, ,
\end{align}
the introduction of the cosmological constant corresponds to the shift of the value of the potential $V(\phi)$. 
Therefore, if we consider the potential, which can be negative even if it is bounded below, we may obtain an NMO. 
Furthermore $F(R)$ gravity can be rewritten in a scalar-tensor form in \eqref{scalartensor}, therefore if we inversely rewrite the scalar-tensor theory with $V(\phi)$, which can be negative, in the form of $F(R)$ gravity, we may obtain an $F(R)$-gravity theory, which may admit NMOs. 

\subsection{Two-scalar model}\label{TwSclr}

In this section, we consider four-dimensional Einstein's gravity coupled with two scalar fields $\phi$ and $\chi$, proposed in Ref.~\cite{Nojiri:2020blr}, 

\subsubsection{Review of two-scalar model}

Based on Ref.~\cite{Nojiri:2020blr}, we review the two-scalar model, whose action is given by
\begin{align}
\label{I8}
S_{\mathrm{GR} \phi\chi} = \int d^4 x \sqrt{-g} &\, \left[ \frac{R}{2\kappa^2}
 - \frac{1}{2} A (\phi,\chi) \partial_\mu \phi \partial^\mu \phi
 - B (\phi,\chi) \partial_\mu \phi \partial^\mu \chi \right. \nonumber \\
& \left. \qquad - \frac{1}{2} C (\phi,\chi) \partial_\mu \chi \partial^\mu \chi
 - V (\phi,\chi) + \mathcal{L}_\mathrm{matter} \right]\, .
\end{align}
Here $A(\phi,\chi)$, $B(\phi,\chi)$, and $C(\phi,\chi)$ are arbitrary functions, $V(\phi,\chi)$ is the scalar-field potential, 
and $\mathcal{L}_\mathrm{matter}$ is the matter Lagrangian density. 

The variation of the action \eqref{I8} with respect to the metric $g_{\mu\nu}$ gives the following Einstein equations, 
\begin{align}
\label{gb4bD4}
0= &\, \frac{1}{2\kappa^2}\left(- R_{\mu\nu} + \frac{1}{2} g_{\mu\nu} R\right) \nonumber \\
&\, + \frac{1}{2} g_{\mu\nu} \left[
 - \frac{1}{2} A (\phi,\chi) \partial_\rho \phi \partial^\rho \phi
 - B (\phi,\chi) \partial_\rho \phi \partial^\rho \chi
 - \frac{1}{2} C (\phi,\chi) \partial_\rho \chi \partial^\rho \chi - V (\phi,\chi)\right] \nonumber \\
&\, + \frac{1}{2} \left[ A (\phi,\chi) \partial_\mu \phi \partial_\nu \phi
+ B (\phi,\chi) \left( \partial_\mu \phi \partial_\nu \chi
+ \partial_\nu \phi \partial_\mu \chi \right)
+ C (\phi,\chi) \partial_\mu \chi \partial_\nu \chi \right] 
+ \frac{1}{2} T_{\mathrm{matter}\, \mu\nu} \, .
\end{align}
Here $\mu, \nu, \ldots=0, 1, 2, 3$ and $T_{\mathrm{matter}\, \mu\nu}$ is the energy-momentum tensor of matter. 
By the variations of the action \eqref{I8} with respect to the fields $\phi$ and $\chi$, we obtain the following field equations, 
\begin{align}
\label{I10}
0 =&\, \frac{1}{2} A_\phi \partial_\mu \phi \partial^\mu \phi
+ A \nabla^\mu \partial_\mu \phi + A_\chi \partial_\mu \phi \partial^\mu \chi
+ \left( B_\chi - \frac{1}{2} C_\phi \right)\partial_\mu \chi \partial^\mu \chi
+ B \nabla^\mu \partial_\mu \chi - V_\phi \, , \nonumber \\
0 =&\, \left( - \frac{1}{2} A_\chi + B_\phi \right) \partial_\mu \phi \partial^\mu \phi
+ B \nabla^\mu \partial_\mu \phi
+ \frac{1}{2} C_\chi \partial_\mu \chi \partial^\mu \chi + C \nabla^\mu \partial_\mu \chi
+ C_\phi \partial_\mu \phi \partial^\mu \chi - V_\chi \, .
\end{align}
Here $A_\phi=\partial A(\phi,\chi)/\partial \phi$, etc. 
Equations in \eqref{I10} are not independent of the Einstein equations in \eqref{gb4bD4}, but they can be obtained from \eqref{gb4bD4} by using the Bianchi identities (see \cite{Nojiri:2024dde} for the details). 

Assume a general spherically symmetric and time-dependent spacetime, 
\begin{align}
\label{GBiv_time}
ds^2 = - \e^{2\nu (t,r)} dt^2 + \e^{2\lambda (t,r)} dr^2 + r^2 \left( d\theta^2 + \sin^2\vartheta d\varphi^2 \right)\, , 
\end{align}
and the following ansatz 
\begin{align}
\label{TSBH1}
\phi=t\, , \quad \chi=r\, ,
\end{align}
without any loss of generality (see \cite{Nojiri:2020blr, Nojiri:2023dvf, Nojiri:2023zlp, Elizalde:2023rds, Nojiri:2023ztz}). 

By using the model~\eqref{I8}, we construct models realising an arbitrarily given spherically symmetric geometry expressed by the metric~\eqref{GBiv_time}. 
In the constructed model, however, the functions $A$ and/or $C$ often become negative, which tells that $\phi$ and/or $\chi$ become ghosts and therefore the model is physically irrelevant. 
The ghosts can be eliminated by imposing constraints by using the Lagrange multiplier fields $\lambda_\phi$ and $\lambda_\chi$. 
We modify the action \eqref{I8} as $S_{\mathrm{GR} \phi\chi} \to S_{\mathrm{GR} \phi\chi} + S_\lambda$. 
Here, the additional term $S_\lambda$ is given by
\begin{align}
\label{lambda1}
S_\lambda = \int d^4 x \sqrt{-g} \left[ \lambda_\phi \left( \e^{-2\nu(t=\phi, r=\chi)} \partial_\mu \phi \partial^\mu \phi + 1 \right)
+ \lambda_\chi \left( \e^{-2\lambda(t=\phi, r=\chi)} \partial_\mu \chi \partial^\mu \chi - 1 \right) \right] \, .
\end{align}
The variations of the term $S_\lambda$ with respect to $\lambda_\phi$ and $\lambda_\chi$ give the following constraints,
\begin{align}
\label{lambda2}
0 = \e^{-2\nu(t=\phi, r=\chi)} \partial_\mu \phi \partial^\mu \phi + 1 \, , \quad
0 = \e^{-2\lambda(t=\phi, r=\chi)} \partial_\mu \chi \partial^\mu \chi - 1 \, .
\end{align}
We should note that the constraints~\eqref{lambda2} are consistent with the assumption~\eqref{TSBH1}.
The constraints~\eqref{lambda2} surely eliminate the ghosts (see \cite{Nojiri:2024dde}). 

In the model given by $S_{\mathrm{GR} \phi\chi} + S_\lambda$, the only propagating mode is a massless spin two mode corresponding to the standard gravitational wave. 
Due to the constraints~\eqref{lambda2}, the scalar fields $\phi$ and $\chi$ do not propagate. 
The Lagrange multiplier fields $\lambda_\phi$ and $\lambda_\chi$ do not propagate, either. 
Although non-trivial energy density and pressure appear, there is no sound, which is a propagating scalar mode. 
In this sense, the fluids expressed by the scalar fields $\phi$, $\chi$, $\lambda_\phi$ and $\lambda_\chi$ are frozen, and any assumed solution becomes stable. 

As shown in Refs.~\cite{Nojiri:2023dvf, Nojiri:2023zlp, Elizalde:2023rds, Nojiri:2023ztz}, even in the model given by the modified action $S_{\mathrm{GR} \phi\chi} + S_\lambda$, $\lambda_\phi=\lambda_\chi=0$ consistently appear as a solution. 
This tells that any solution of Eqs.~\eqref{gb4bD4} and \eqref{I10} corresponding to the original action~\eqref{I8} is a solution for the modified model given by the action $S_{\mathrm{GR} \phi\chi} + S_\lambda$. 
We should also note that there is an asymptotically flat solution in the model given by $S_{\mathrm{GR} \phi\chi} + S_\lambda$, any solution of the Einstein equation, even if we include matter, is a solution of the model (see \cite{Nojiri:2023ztz}). 

\subsubsection{Construction of models which realise a negative mass object}\label{sec:Reconstruction}

Let us try to construct a model which has a solution realising an NMO expressed by the metric functions $\e^{2\nu(t,r)}$ and $\e^{2\lambda(t,r)}$ given in Eq.~\eqref{GBiv_time}. 

The $(t,t)$, $(r,r)$, $(\vartheta,\vartheta)$, and $(t,r)$ components in Eqs.~\eqref{gb4bD4} are expressed as 
\begin{align}
\label{TSBH2}
\frac{\e^{-2\lambda + 2\nu}}{\kappa^2} \left( \frac{2\lambda'}{r} + \frac{\e^{2\lambda} - 1}{r^2} \right)
=&\, - \e^{2\nu} \left( - \frac{A}{2} \e^{-2\nu} - \frac{C}{2} \e^{-2\lambda} - V \right) + \e^{2\nu} \rho \, , \nonumber \\
\frac{1}{\kappa^2} \left( \frac{2\nu'}{r} - \frac{\e^{2\lambda} - 1}{r^2} \right)
=&\, \e^{2\lambda} \left( \frac{A}{2} \e^{-2\nu} + \frac{C}{2} \e^{-2\lambda} - V \right) + \e^{2\lambda} p \, , \nonumber \\
\frac{1}{\kappa^2} \left\{ - r^2 \e^{-2 \nu} \left[ \ddot\lambda \right. \right. & + \left. \left. \left( \dot\lambda - \dot\nu \right) \dot\lambda \right]
+ \e^{-2\lambda}\left( r \left(\nu' - \lambda' \right) + r^2 \nu'' + r^2 \left( \nu' - \lambda' \right) \nu' \right) \right\} \nonumber \\
=&\, r^2 \left( \frac{A}{2} \e^{-2\nu} - \frac{C}{2} \e^{-2\lambda} - V \right) + r^2 p \, , \nonumber \\
\frac{2\dot\lambda}{\kappa^2 r} =&\, B \, .
\end{align}
Here, the dot `$\dot\ $' and prime `$\ `$' denote differentiation with respect to the time coordinate $t$ and radial coordinate $r$, respectively. 
We also assume that the matter is a perfect fluid with the energy density $\rho$ and the pressure $p$, 
\begin{align}
\label{FRk2}
T_{\mathrm{matter}\, tt} =-g_{tt}\rho\ ,\quad T_{\mathrm{matter}\, ij}=p\, g_{ij}\, ,
\end{align} 
where $i, j= r,\vartheta, \varphi$.
The equations~\eqref{TSBH2} are algebraically solved with respect to $A$, $B$, $C$, and $V$ below, 
\begin{align}
\label{ABCV}
A=& \frac{\e^{2\nu}}{\kappa^2} \left\{ - \e^{-2 \nu} \left[ \ddot\lambda + \left( \dot\lambda - \dot\nu \right) \dot\lambda \right]
+ \e^{-2\lambda}\left[ \frac{\nu' + \lambda'}{r} + \nu'' + \left( \nu' - \lambda' \right) \nu' + \frac{\e^{2\lambda} - 1}{r^2}\right] \right\}
 - \e^{2\nu} \left( \rho + p \right) \, , \nonumber \\
B=&\, \frac{2\dot\lambda}{\kappa^2 r} \, , \nonumber \\
C=&\, \frac{\e^{2\lambda}}{\kappa^2} \left\{ \e^{-2 \nu} \left[ \ddot\lambda + \left( \dot\lambda - \dot\nu \right) \dot\lambda \right]
 - \e^{-2\lambda}\left[ - \frac{\nu' + \lambda'}{r} + \nu'' + \left( \nu' - \lambda' \right) \nu' + \frac{\e^{2\lambda} - 1}{r^2}\right] \right\} \, , \nonumber \\
V=& \frac{\e^{-2\lambda}}{\kappa^2} \left( \frac{\lambda' - \nu'}{r} + \frac{\e^{2\lambda} - 1}{r^2} \right) - \frac{1}{2} \left( \rho - p \right) \, .
\end{align}
This shows that one can construct a model which realises the spacetime defined by the metric \eqref{GBiv_time} by finding $(t,r)$-dependence of $\rho$ and $p$ and 
by replacing $(t,r)$ in Eq.~\eqref{ABCV} with $(\phi,\chi)$.

The metric of the spacetime describing the Hayward black hole~\cite{Hayward:2005gi}, which is a static and regular black hole without curvature singularity, is given by
\begin{align}
\label{Hayward}
\e^{2\nu}=\e^{-2\lambda} = 1 - \frac{2Mr^2}{r^3 + 2M\lambda^2} \, .
\end{align}
Here $M>0$ is the Arnowitt-Deser-Misner (ADM) mass. 

As an NMO, we consider the following 
\begin{align}
\label{alaHayward}
\e^{2\nu}=\e^{-2\lambda} = 1 + \frac{2Mr^2}{r^3 + 2M\lambda^2} \, .
\end{align}
We should note that the ADM mass is negative $-M$. 
Due to the anti-gravity generated by negative mass, 
a photon sphere does not appear. 
The object works as a gravitational concave lens. 

We should also note that due to the constraints~\eqref{lambda2}, the solution~\eqref{alaHayward} is stable because the scalar fields are frozen. 

As discussed in \cite{Nojiri:2023ztz}, we now show that there are solutions other than the assumed spacetime as in \eqref{GBiv_time}. 
We now assume that the given spacetime is asymptotically flat, asymptotically de Sitter (dS) or anti-de Sitter (adS). 
Under the assumption, any solution in Einstein's gravity is also a solution of the reconstructed model, even if we include matter.
First, we should note that under the assumption that the spacetime is asymptotically flat, dS, or adS when $t \rightarrow \pm \infty$ or $r \rightarrow \infty$, the static or eternal flat, dS, and adS spacetimes are also solutions automatically. 
This is because by considering the limit $t_0 \rightarrow \pm \infty$ after shifting the time coordinate $t \rightarrow t_0+t$ or the limit $r_0 \rightarrow \infty$ after shifting the radial coordinate $r \rightarrow r_0+r$, the obtained solution describes flat, dS, or adS spacetime. 

Even in the case of the black hole solution in Einstein's gravity, it is difficult to find an exact solution describing two or more black holes. 
We believe, however, that such solutions should exist because the spacetime with infinitely separated black holes should be a solution of Einstein's gravity. 
Similarly, even in our model, the model could describe the spacetime with several NMOs. 

\subsection{Dynamical creation of Negative Mass Object?}

For the model with two scalar fields $\phi$ and $\chi$ in \eqref{I8}, one can construct a time-dependent spacetime. 
Therefore, we may construct a model where an NMO is created by the accretion of the scalar fields. 
An example is given by 
\begin{align}
\label{alaHaywarddynmc}
\e^{2\nu}=\e^{-2\lambda} = 1 + \frac{2Mr^2}{\left(1+ \e^{-\frac{t}{t_0}}\right)\left(r^3 + 2M\lambda^2\right)} \, .
\end{align}
Here $t_0$ is a positive constant. 
When $t\to -\infty$, the metric reduces to that of the flat Minkowski spacetime, $\e^{2\nu}=\e^{-2\lambda} \to 1$. 
On the other hand, in the limit of $t\to +\infty$, the metric coincides with the NMO~\eqref{alaHayward}. 
Therefore, the metric given by \eqref{alaHaywarddynmc} surely expresses the creation of the NMO. 

Note that for the metric~\eqref{GBiv_time}, the $(r,t)$-component of the Ricci tensor $R_{\mu\nu}$, which also gives the $(r,t)$ components of the total energy-momentum tensor $T_{\mu\nu}$, is given by 
\begin{align}
\label{Rrt}
R_{rt} =R_{tr} = \kappa^2 T_{rt} = \kappa^2 T_{tr} = \frac{2\dot \lambda}{r}\, .
\end{align}
Using \eqref{alaHaywarddynmc}, one get
\begin{align}
\label{Trt}
T_{rt}=T_{tr} = - \frac{2Mr^2\e^{-\frac{t}{t_0}}}{\kappa^2 t_0 r\left(1+ \e^{-\frac{t}{t_0}}\right)^2\left(r^3 + 2M\lambda^2\right)
\left\{ 1 + \frac{2Mr^2}{\left(1+ \e^{-\frac{t}{t_0}}\right)\left(r^3 + 2M\lambda^2\right)}\right\}^2} \, ,
\end{align}
which is negative. 

The flow of the energy density is expressed by 
\begin{align}
\label{Trt2}
T_r^{\ t}=T^t_{\ r} = \frac{2Mr^2\e^{-\frac{t}{t_0}}}{\kappa^2 t_0 r\left(1+ \e^{-\frac{t}{t_0}}\right)^2\left(r^3 + 2M\lambda^2\right)
\left\{ 1 + \frac{2Mr^2}{\left(1+ \e^{-\frac{t}{t_0}}\right)\left(r^3 + 2M\lambda^2\right)}\right\}}\, ,
\end{align}
which is positive. 
Therefore, there appears the positive energy flow to the radial direction, which is the inverse process of the accretion of matter creating the usual stars. 

When the contributions from matter are neglected, the last expression in \eqref{ABCV} gives the following form of the potential $V(\phi,\chi)$, 
\begin{align}
\label{ABCV2}
V(\phi,\chi) = - \frac{8M^2\lambda^2 \chi}{\kappa^2\chi\left(1+ \e^{-\frac{\phi}{t_0}}\right)\left(\chi^3 + 2M\lambda^2\right)^2} \, .
\end{align}
With respect to $\phi$, the potential has a maximum when $\phi\to-\infty$, $V(\phi=-\infty,\chi)$, and the potential has a minimum when $\phi\to+\infty$, $V(\phi=-\infty,\chi)=- \frac{8M^2\lambda^2 \chi}{\kappa^2\chi\left(\chi^3 + 2M\lambda^2\right)^2}$, 
Therefore, the NMO is created by the rolling down of the scalar field $\phi$ along the potential, and the details of the creation depend on the initial conditions of the scalar fields. 

The above proposed process of the creation of the NMOs is a kind of decay of the false vacuum corresponding to $\phi=-\infty$ and the true vacuum corresponding to $\phi=\infty$ as in the Higgs mechanism. 
In the Higgs mechanism, there appear domain walls between the false vacuum and the true vacuum. 
The domain walls grow up, and the whole universe becomes the true vacuum. 
In our model, however, because the scalar fields are frozen due to the constraints~\eqref{lambda2}, they do not grow. 
Eventually, other scenarios to generate NMOs may be proposed, too.


\subsection{General spherically symmetric and static solution of scalar-Einstein-Gauss-Bonnet gravity}\label{SectionIIIB}

In this section, in the framework of the scalar-Einstein-Gauss-Bonnet gravity, the realisation of an NMO is also achieved. 
We used the formulation refined from that in \cite{Nojiri:2023qgd}. 

Let us write the action of the scalar-Einstein-Gauss-Bonnet gravity in the following form, 
\begin{align}
\label{g2}
\mathcal{S}=\int d^N x \sqrt{-g}\left\{ \frac{1}{2\kappa^2}R - \frac{1}{2} A(\xi) \partial_\mu \xi \partial^\mu \xi - V(\xi)+ f(\xi) \mathcal{G} \right\}\, .
\end{align}
Here $\xi$ is the scalar field, $A(\xi)$ and $f(\xi)$ are functions of $\xi$ and $V(\xi)$ is the potential for $\xi$. 
In \eqref{g2}, $\mathcal{G}$ is the Gauss-Bonnet invariant defined by
\begin{align}
\label{eq:GB}
\mathcal{G} = R^2-4R_{\alpha \beta}R^{\alpha \beta}+R_{\alpha \beta \rho \sigma}R^{\alpha \beta \rho \sigma}\, ,
\end{align}
Assume that matter does not directly couple with the scalar field $\xi$. 

The variation of the action~\eqref{g2} with respect to the scalar field $\xi$ yields the following equation,
\begin{align}
\label{g3}
A(\xi) \nabla^2 \xi - \frac{1}{2} A'(\xi) \partial_\mu \xi \partial^\mu \xi - V'(\xi) + f'(\xi)\mathcal{G}=0\, .
\end{align}
On the other hand, by the variation of the action \eqref{g2} with respect to the metric $g_{\mu\nu}$, we obtain the following equations,
\begin{align}
\label{gb4bD4B}
0= & \frac{1}{2\kappa^2}\left(- R^{\mu\nu} + \frac{1}{2} g^{\mu\nu} R\right)
+ A(\xi) \left(\frac{1}{2} \partial^\mu \xi \partial^\nu \xi
 - \frac{1}{4}g^{\mu\nu} \partial_\rho \xi \partial^\rho \xi \right)
 - \frac{1}{2} g^{\mu\nu}V(\xi) \nonumber \\
& + 2 \left( \nabla^\mu \nabla^\nu f(\xi)\right)R
 - 2 g^{\mu\nu} \left( \nabla^2f(\xi)\right)R
 - 4 \left( \nabla_\rho \nabla^\mu f(\xi)\right)R^{\nu\rho}
 - 4 \left( \nabla_\rho \nabla^\nu f(\xi)\right)R^{\mu\rho} \nonumber \\
& + 4 \left( \nabla^2 f(\xi) \right)R^{\mu\nu}
+ 4g^{\mu\nu} \left( \nabla_{\rho} \nabla_\sigma f(\xi) \right) R^{\rho\sigma}
- 4 \left(\nabla_\rho \nabla_\sigma f(\xi) \right) R^{\mu\rho\nu\sigma}.
\end{align}
The scalar field equation~\eqref{g3} can be obtained from Eq.~\eqref{gb4bD4B} by using the Bianchi identity. 

In the following, consider the spherically symmetric and static spacetime, where the metric is given by $\e^{2\nu}=\e^{-2\lambda} = a(r)$ in \eqref{GBiv}. 
Then the equations in Eq.~\eqref{gb4bD4B} are,
\begin{align}
\label{Eq2tt}
0=&\, \frac{16a\left(1 -a\right) f'' + \left\{ 8\left( 1-3a \right)f' +2r \right\} a'_1
+2a +r^2 A(\xi) \xi'^2 a -2+2r^2V}{4r^2} \, , \\
\label{Eq2rr}
0=&\, \frac{2\left(4 \left(1 -3a \right) f' +r \right)a'+2a -r^2 a A(\xi) \xi'^2 -2 +2V r^2 }{4 r^2 }\, , \\
\label{Eq2pp}
0=&\, \frac{\left(r - 8 f' a \right)a'' -8 f''a a' - 8 f' {a'}^2 + 2 a' +r\left(2V+ A(\xi) \xi'^2a \right)}{4r} \, .
\end{align}
Eqs.~\eqref{Eq2tt} and \eqref{Eq2rr} correspond to the $(t,t)$-component and $(r,r)$ -componet of Eq.~\eqref{gb4bD4B}, respectively.
On the other hand, from the $(\theta,\theta)$ and $(\varphi,\varphi)$-components, we obtain the identical equation~\eqref{Eq2pp}.

By combining Eq.~\eqref{Eq2tt} and Eq.~\eqref{Eq2rr}, one finds
\begin{align}
\label{V2}
V = \frac{1-4 a \left( 1-a \right) f'' - \left[ 4a' \left( 1-3a \right) f'+r \right] a'-a }{r^2} \, .
\end{align}
As in \eqref{TSBH1}, if we also assume 
\begin{align}
\label{xir}
\xi=r\, ,
\end{align}
by using Eq.~\eqref{Eq2tt} and Eq.~\eqref{Eq2rr}, we find
\begin{align}
\label{xi3}
A(\xi) =\frac{8}{r^2}\left( a-1 \right)f'' \, .
\end{align}
Finally, by using Eq.~\eqref{Eq2tt} and Eq.~\eqref{Eq2pp}, we obtain 
\begin{align}
\label{f}
0= 8a \left( 2a-a'r-2 \right)f''+a''r \left( r-8f'a \right)-8rf'a'^2+8 \left( 3a-1 \right)f'a'+2 \left( 1-a \right) \, .
\end{align}
Eq.~\eqref{xi3} shows if $\left( a-1 \right)f''<0$, $\xi$ becomes a ghost. 
In order to solve this problem, as in \eqref{lambda1}, we include the term $S_\lambda$ with the Lagrange multiplier field $\lambda$, 
\begin{align}
\label{lambda1GB}
S_\lambda = \int d^4 x \sqrt{-g} \lambda_\chi \left( a(r) \partial_\mu \xi \partial^\mu \xi - 1 \right) \, .
\end{align}
Then the variations of the term $S_\lambda$ in \eqref{lambda1GB} with respect to $\lambda$ give the following constraint, 
\begin{align}
\label{lambda2GB}
0 = a(r) \partial_\mu \xi \partial^\mu \xi - 1 \, ,
\end{align}
which excludes the ghost and any instabilities associated with the scalar field $\xi$. 

We should note that Eq.~\eqref{f} can be trivially integrated,
\begin{align}
\label{f1}
f(r)=-\frac{1}{8}\int \left( \int \frac{\e^{\int \frac{a' -3 a a' +r{a'}^2+ra a'' }{a \left( 2-2\,a + a' r \right)}dr}
\left(2 a-2 - a'' r^2 \right) }{ a\left( 2-2a + a' r \right)}dr -8 c_0 \right) { \e^{-\int \frac{a' -3 a a' +r {a'}^2+ra a'' }{a \left( 2-2 a + a' r \right) }dr}} dr+c_1\,,
\end{align}
where $c_0$ and $c_1$ are constants of integration.

When $a=a(r)$ is given, we find the $r$-dependence of $f$, i.e., $f=f(r)$ by using \eqref{f1}.
Then by using \eqref{V2}, one gets the $r$-dependence of the potential $V=V(r)$ and furthermore by using \ref{xi3}, we also obtain $A(\xi)=A(r)$ as a function of $r$. 
Due to \eqref{xir}, $\xi=r$, by replacing $r$ in the expression of $f(r)$, $V(r)$, and $A(r)$ with $\xi$, we obtain the model to realise the metric in \eqref{GBiv} with $\e^{2\nu}=\e^{-2\lambda} = a(r)$.
Then we may consider the metric~\eqref{alaHayward} a la that of the Hayward black hole~\eqref{Hayward}. 
We should also note that due to the constraints~\eqref{lambda2GB}, the solution~\eqref{alaHayward} is stable, again, because the scalar field $\xi$ is frozen. 

\section{Orbit of Photon}\label{Phtn}

Some models realising the metric given by \eqref{alaHayward} have been obtained. 
Since the ADM mass is negative in the spacetime given by the metric~\eqref{alaHayward}, the anti-gravity could work as a gravitational concave lens. 
In this section, we investigate the orbit of the photon in the NMO geometry by assuming that the orbit could be given by a null orbit. 

The motion of the photon is described by the following Lagrangian, 
\begin{align}
\label{ph1g}
\mathcal{L}= \frac{1}{2} g_{\mu\nu} \dot q^\mu \dot q^\nu = \frac{1}{2} \left( - \e^{2\nu} {\dot t}^2 + \e^{2\lambda} {\dot r}^2 + r^2 {\dot\theta}^2 + r^2 \sin^2 \theta {\dot\varphi}^2 \right) \, .
\end{align}
Here, the dot ``$\dot\ $'' expresses the derivative with respect to the affine parameter. 
Because the Lagrangian $\mathcal{L}$ does not depend on the $t$ and $\varphi$, 
there exist conserved quantities corresponding to energy $E$ and angular momentum $L$, 
\begin{align}
\label{phEg}
E \equiv&\, \frac{\partial \mathcal{L}}{\partial \dot t} = - \e^{2\nu} \dot t \, , \\
\label{phMg}
L \equiv&\, \frac{\partial V}{\partial\dot\varphi}= r^2 \sin^2 \theta \dot\varphi \, . 
\end{align}
The total energy $\mathcal{E}$ of the system should be also conserved, 
\begin{align}
\label{totalEg}
\mathcal{E} \equiv \mathcal{L} - \dot t \frac{\partial \mathcal{L}}{\partial \dot t} - \dot r \frac{\partial \mathcal{L}}{\partial \dot r} 
 - \dot\theta \frac{\partial \mathcal{L}}{\partial \dot\theta} - \dot\varphi \frac{\partial \mathcal{L}}{\partial \dot\varphi} = \mathcal{L} \, . 
\end{align}
In the case of a photon, whose geodesic is null, we require $\mathcal{E}=\mathcal{L}=0$. 

Without loss of generality, we consider the orbit on the equatorial plane with $\theta=\frac{\pi}{2}$. 
Then the condition $\mathcal{E}=\mathcal{L}=0$ gives, 
\begin{align}
\label{geo1g}
0= - \frac{E^2}{2} + \frac{1}{2} {\dot r}^2 + \frac{L^2 a(r)}{2r^2} \, .
\end{align}
Here, we considered the case $\e^{2\nu}=\e^{-2\lambda}=a(r)$ as in \eqref{alaHayward}. 

This system is analogous to the classical dynamical system with potential $U(r)$, 
\begin{align}
\label{geo2g}
0 =\frac{1}{2} {\dot r}^2 + U(r)\, , \quad U(r) \equiv \frac{L^2 a(r)}{2r^2} - \frac{E^2}{2} \, .
\end{align}
If there is a circular orbit, where $\dot r=0$, the equations $U(r)= U'(r)=0$ must have a solution. 
We find, however, 
\begin{align}
\label{phg1}
U(r) =&\, \frac{L^2}{2r^2} \left( 1 + \frac{2Mr^2}{r^3 + 2M\lambda^2} \right) - \frac{E^2}{2} \, , \\
\label{phg2}
U'(r) =&\, - \frac{L^2}{r} \left( \frac{1}{r^2} + \frac{3Mr^3}{ \left(r^3 + 2M\lambda^2 \right)^2} \right) <0 \, , 
\end{align}
and Eq.~\eqref{phg2} tells that there is no solution. 
As expected, there is no circular orbit of a photon, nor a photon sphere. 

As in classical mechanics, the solution of \eqref{geo2g} is given by 
\begin{align}
\label{orbit}
t = \int \frac{dr}{\sqrt{-2U(r)}} = \int \frac{dr}{\sqrt{E^2 - \frac{L^2}{r^2} \left( 1 + \frac{2Mr^2}{r^3 + 2M\lambda^2} \right)}} \, ,
\end{align}
or because Eq.~\eqref{phMg} with $\theta=\frac{\pi}{2}$ tells $\frac{d\varphi}{dt} = \frac{L}{r^2}$, we rewrite \eqref{geo2g} with \eqref{phg1} in the following form, 
\begin{align}
\label{geo2gB}
0 =\frac{L^2}{2r^4} \left(\frac{dr}{d\varphi} \right)^2 + \frac{L^2}{2r^2} \left( 1 + \frac{2Mr^2}{r^3 + 2M\lambda^2} \right) - \frac{E^2}{2} \, ,
\end{align}
which can be integrated to give, 
\begin{align}
\label{orbitph}
\varphi = \int \frac{dr}{r^2 \sqrt{\frac{E^2}{L^2} - \frac{1}{r^2} \left( 1 + \frac{2Mr^2}{r^3 + 2M\lambda^2} \right)}} \, , 
\end{align}
which gives the shape of the orbit. 
An asymptotically hyperbolic curve could give the orbit. 
Therefore, the NMO can be regarded as a gravitational concave lens.

In the limit of $\lambda\to 0$, the geometry becomes the Schwarzschild spacetime with negative mass, and there appears a naked singularity. 
In the region far from the singularity, the effects of the singularity could be neglected. 
In the limit, the expression in \eqref{orbitph} is approximated as, 
\begin{align}
\label{orbitphax}
\varphi = \int \frac{dr}{r^2 \sqrt{\frac{E^2}{L^2} - \frac{1}{r^2} \left( 1 + \frac{2M}{r} \right)}} \, . 
\end{align}
Although the qualitative structure is clear, the integration in \eqref{orbitphax} is not given by an elementary function. 
A special case where we can integrate is given by the case, 
\begin{align}
\label{cdint}
\frac{E^2}{L^2} = \frac{1}{27 M^2}\, .
\end{align}
In the case of \eqref{cdint}, Eq.~\eqref{orbitphax} has the following form, 
\begin{align}
\label{orbitphax2}
\varphi = \int \frac{3\sqrt{3} M dr}{\left( r + 3M \right) \sqrt{\left(r - 3 M\right)^2 - 9 M^2}} \, . 
\end{align}
Then by changing the variable by $r=3M \cosh \zeta$ $\left(dr = 3M \sinh \zeta d\zeta\right)$, we obtain the following expression, 
\begin{align}
\label{orbitphax3}
\varphi = \frac{\sqrt{3}}{2} \tanh \frac{\zeta}{2} \, . 
\end{align}
Here, the constant of the integration is absorbed into the redefinition of $\varphi$. 
This expression \eqref{orbitphax3} gives the shape of the orbit $\varphi=\varphi(r)$ or $r=r(\varphi)$, 
\begin{align}
\frac{4\varphi^2}{3} = \tanh^2 \frac{\zeta}{2} = \frac{ \cosh \zeta - 1}{\cosh \zeta + 1} = \frac{r-3M}{r+3M} \, , \quad 
r= 3M \frac{ 1 + \frac{4\varphi^2}{3} }{1 - \frac{4\varphi^2}{3} }\, .
\end{align}
The minimum of $r$ is given by $r=3M$ when $\zeta=\varphi=0$, which, maybe accidentally, coincides with the radius of the photon sphere of the Schwarzschild black hole. 

When $r\to \infty$, we find $\varphi\to \pm \frac{\sqrt{3}}{2}$ and therefore the scattering angle $\Phi$ is given by 
\begin{align}
\label{sa}
\Phi = \pi - \sqrt{3} \, .
\end{align}
Therefore, the NMO surely works as a gravitational concave lens. 
If the object exists in front of the light source(s), for example, a small NMO in front of the sun, the object could be observed as a dark spot, although we need some numerical calculations to obtain the exact image. 
Of course, this consideration is completely classical. 

\section{Behaviour for massive particle}\label{mssv}

Instead of a photon, we may consider massive particles. 
For the massive particles, we may consider the Newton limit, where
\begin{itemize}
\item Gravity is weak, that is, $\e^{2\nu}\sim \e^{2\lambda} \sim 1$. 
\item The spacetime is static or quasi-static. 
\item The speed $v$ of the particle is much smaller than the speed $c$ of light, $\frac{v}{c}\ll 1$ (only here, we explicitly write the speed of light as $c$). 
\end{itemize}
Then the Newton potential $\Psi$ is given by 
\begin{align}
\label{NwtnPt}
\e^{2\nu}=1 + 2\Psi\, .
\end{align}
In the case that the spacetime is static and spherically symmetric, we find that $\Psi$ only depends on the radial coordinate $r$, $\Psi=\Psi(r)$. 
If $\Psi(r)$ is an increasing function of $r$, the gravity is attractive but if $\Psi(r)$ is a decreasing function of $r$, the gravity is repulsive, that is, the anti-gravity appears. 

In the case of the Hayward black hole~\eqref{Hayward}, the Newton potential is given by, 
\begin{align}
\label{HaywardNwtn0}
\Psi(r) = - \frac{Mr^2}{r^3 + 2M\lambda^2} \, .
\end{align}
When $r$ is large, $\Psi(r)$ behaves as 
\begin{align}
\label{HaywardNwtn}
\Psi(r) \sim - \frac{M}{r} \, .
\end{align}
Therefore as long as the ADM mass $M$ is positive, the Newton potential is attractive. 
However, 
\begin{align}
\label{HaywardNwtn2}
\Psi'(r) 
= - \frac{2Mr}{r^3 + 2M\lambda^2} + \frac{3Mr^4}{\left(r^3 + 2M\lambda^2\right)^2}
= \frac{Mr \left(r^3 - 4M\lambda^2\right)}{\left(r^3 + 2M\lambda^2\right)^2} \, ,
\end{align}
the repulsive force appears when $r<\left( 4M\lambda^2 \right)^\frac{1}{3}$. 

On the other hand, for an NMO in \eqref{alaHayward}, we find 
\begin{align}
\label{alaHaywardB}
\e^{2\nu}=\frac{Mr^2}{r^3 + 2M\lambda^2} \, , 
\end{align}
and therefore 
\begin{align}
\label{alaHaywardNwtn}
\Psi(r) = \frac{Mr^2}{r^3 + 2M\lambda^2} \, .
\end{align}
Then the repulsive force appears when $r>\left( 4M\lambda^2 \right)^\frac{1}{3}$ and the attractive force appears when $r<\left( 4M\lambda^2 \right)^\frac{1}{3}$. 

Since the NMO gives the repulsive force to the usual positive massive object, the NMO suffers the repulsive force due to Newton's third law of motion. 
We should note, however, that the inertial mass of the NMO could be negative because we require the equivalence principle. 
Therefore, the repulsive force effectively works as an attractive force and the NMOs gather to the standard galaxies. 
This can be understood by assuming that the Newton approximation of gravity is valid and Newton's second law of motion works. 
Then we obtain, 
\begin{align}
\label{2ndlw}
m_\mathrm{inertia} \ddot{\bm r} = - m_\mathrm{gravity} \nabla \Psi(r)\, .
\end{align}
Here $m_\mathrm{inertia}$ is the inertial mass and $m_\mathrm{gravity}$ is the gravitational mass. 
The equivalence principle, $m_\mathrm{inertia} = m_\mathrm{gravity}=m$, the second law~\eqref{2ndlw} reduces to the following form, 
\begin{align}
\label{2ndlw2}
\ddot{\bm r} = - \nabla \Psi(r)\, .
\end{align}
The expression~\eqref{2ndlw2} does not depend on the signature of the mass $m$. 
This indicates that the Newton force of gravity is attractive for the object with usual positive mass, the force acts as an attractive force even for the NMO. 
On the other hand, the Newton force of gravity is repulsive for the object with usual positive mass, the force acts as a repulsive force even for the NMO. 
That is, the force between two NMOs is repulsive. 
The above results are consistent with those in \cite{Manfredi:2026qoz}. 

The above results are consistent if we consider the geodesic. 
In order to investigate the behaviour of the massive particle behind the Newton limit, we need to use the geodesic. 
The Lagrangian for the geodesic is given by \eqref{ph1g} and the conservation laws are also given by \eqref{phEg} and \eqref{phMg} but the total energy $\mathcal{E}$ in \eqref{totalEg} is modified to be $\mathcal{E}=\mathcal{L}=-\frac{m^2}{2}$. 
Here $m$ is the mass of the particle. 
Then instead of \eqref{geo2g}, we obtain 
\begin{align}
\label{geo2gmp}
0 =\frac{1}{2} {\dot r}^2 + U(r)\, , \quad U(r) \equiv \frac{L^2 a(r)}{2r^2} - \frac{E^2}{2} + \frac{m^2 a(r)}{2}\, .
\end{align}
If there is a solution satisfying $U(r)=U'(r)=0$, 
one gets
\begin{align}
\label{phg1mss}
U(r) =&\, \frac{L^2}{2r^2} \left( 1 + \frac{2Mr^2}{r^3 + 2M\lambda^2} \right) - \frac{E^2}{2} + \frac{m^2}{2}\left( 1 + \frac{2Mr^2}{r^3 + 2M\lambda^2} \right) \, , \\
\label{phg2mss}
U'(r) =&\, \frac{- L^2 \left( r^6 + 4M\lambda^2 r^3 + 4 M^2 \lambda^4 + 3M r^5 \right) + m^2 \left( - M r^7 + 4 M^2 \lambda^2 r^4 \right)
}{r^3 \left( r^3 + 2M\lambda^2 \right)^2} \, .
\end{align}
Because $r^6 + 4M\lambda^2 r^3 + 4 M^2 \lambda^4 + 3M r^5>0$ and $- M r^7 + 4 M^2 \lambda^2 r^4>0$ when $r<\left( 4M\lambda^2 \right)^\frac{1}{3}$, 
if $m^2$ is large enough, there is a solution for $U'(r)=0$ where $r<\left( 4M\lambda^2 \right)^\frac{1}{3}$. 
The condition $r<\left( 4M\lambda^2 \right)^\frac{1}{3}$ is consistently identical to the condition that the Newton potential~\eqref{alaHaywardNwtn} becomes attractive. 
Inversely, if $r>\left( 4M\lambda^2 \right)^\frac{1}{3}$, the massive particle always receives a repulsive force from the NMO. 

We should note that the equation~\eqref{geo2gmp} of the geodesic does not depend on the signature of $m$. 
(We should distinguish it from tachyon, where $m^2$ is negative. )
This is consistent with the results that even the force between two NMOs is also repulsive in the Newton approximation in \eqref{2ndlw} and that the photon with vanishing mass receives the repulsive force in the last section. 

Then if there is a galaxy or massive object, the NMOs will be drawn to the galaxy or the object. 
On the other hand, the repulsive force works between the NMOs. 
This situation could be similar to an atom, where the nucleus attracts electrons, but the force between the electrons is repulsive. 
In the case of the atom, the positive electric charge is screened by the cloud of electrons. 
If we consider the system composed of a positively massive object like a galaxy with NMOs, the mass of the positively massive object is screened by the cloud of the NMOs. 
Such a system might become gravitationally invisible, but for the light, the positively massive object works as a convex lens, and the NMOs work as concave lenses. 
Therefore, such a system might be observed optically through a telescope. 

\section{Undergraduate mechanics of negative mass object(s)}\label{udrgrdtmch}

We may consider the mechanics on the undergraduate level for the system, including NMO(s). 
Here as an approximation, we use Newton's law of mechanics and Newton's law of gravity. 

We assume that there are two point masses with the position vectors $\bm{r}_1$ and $\bm{r}_2$ and the masses $m_1$ and $m_2$. 
Then the equations of motion are given by 
\begin{align}
\label{eom}
m_1 \ddot{\bm{r}}_1 = - \frac{G m_1 m_2 \left( \bm{r}_1 - \bm{r}_2 \right)}{\left| \bm{r}_1 - \bm{r}_2 \right|^3}\, , \quad
m_2 \ddot{\bm{r}}_2 = \frac{G m_1 m_2 \left( \bm{r}_1 - \bm{r}_2 \right)}{\left| \bm{r}_1 - \bm{r}_2 \right|^3}\, .
\end{align}
Then 
\begin{align}
\label{mmtm}
0= m_1 \ddot{\bm{r}}_1 + m_2 \ddot{\bm{r}}_2
= \frac{d}{dt} \left( m_1 \dot{\bm{r}}_1 + m_2 \dot{\bm{r}}_2 \right) \, , 
\end{align}
which shows the conservation of the total momentum as usual. 

First, we consider the case that $m_1 + m_2 \neq 0$. 
We will consider the case that $m_1=m_2$ later, where a strange phenomena could occur. 

We now define the total mass $M$, the reduced mass $\mu$, the position vector of the center of mass $\bm{R}$ and the relative position vector $\bm{r}$, as follows, 
\begin{align}
\label{MmuRr}
M\equiv m_1 + m_2 \, , \quad 
\mu \equiv \frac{m_1 m_2}{m_1 + m_2} \, , \quad 
\bm{R} \equiv \frac{m_1 \bm{r}_1 + m_2 \bm{r}_2}{m_1 + m_2} \, , \quad 
\bm{r} \equiv \bm{r}_2 - \bm{r}_1 \, .
\end{align}
Then we find 
\begin{align}
\label{eom2}
M \ddot{\bm{R}} = 0 \, , \quad 
\mu \ddot{\bm{r}} = - \frac{G \mu M \bm{r}}{\left|\bm{r}\right|^3}\, . 
\end{align}
We may also define a total energy $E$ by 
\begin{align}
\label{tE}
E\equiv&\, \frac{1}{2} m_1 \left| \dot{\bm{r}}_1 \right|^2 + \frac{1}{2} m_2 \left| \dot{\bm{r}}_2 \right|^2 - \frac{Gm_1 m_2}{\left| \bm{r_1} - \bm{r_2} \right|} 
= E_\mathrm{COM} + E_\mathrm{rel}\, , \nonumber \\
&\, E_\mathrm{COM} \equiv \frac{1}{2} M \left| \dot{\bm{R}} \right|^2 \, , \quad 
E_\mathrm{rel} \equiv \frac{1}{2} \mu \left| \dot{\bm{r}} \right|^2 - \frac{G\mu M}{\left|\bm{r}\right|} \, .
\end{align}
Note $E$, $E_\mathrm{COM}$, $E_\mathrm{rel}$ are conserved, respectively. 

The angular momentum $\bm{L}$ with respect to the relative position vector $\bm{r}$, $\bm L = \mu \bm{r} \times \dot{\bm{r}}$ is conserved. 
Then by using the standard procedure in the elementary textbook of mechanics, the last expression in Eq.~\eqref{tE} has the following form, 
\begin{align}
\label{tE3}
E_\mathrm{rel} = \frac{1}{2} \mu {\dot r}^2 + U(r)\, , \quad 
U(r) \equiv \frac{l^2}{2\mu r^2} - \frac{G\mu M}{r} \, .
\end{align}
Here $l\equiv \equiv \left| \bm{L} \right|$. 

By using the standard results of classical mechanics, the centre of mass moves with uniform linear motion. 
If $M>0$, that is, $m_1>-m_2$, the orbit of the relative position vector becomes ellipse or circle when $\frac{E_\mathrm{rel}}{\mu}<0$, a parabola when $\frac{E_\mathrm{rel}}{\mu}=0$, and hyperbolic when $\frac{E_\mathrm{rel}}{\mu}>0$, respectively. 
If $M<0$, the only possible orbit if hyperbolic. 

{\color{red}
When the inertial mass of a point mass is negative, if it suffers any force, the direction of the acceleration of the point mass is opposite to the direction of the force. Therefore, the point mass climbs up the potential. Therefore, if the potential is bounded above, the system becomes stable. 
In the expression of $E_\mathrm{rel}$ in \eqref{tE3}, when the reduced mass $\mu$ is negative, the effective potential is bounded above, and therefore the system becomes stable. 
}

We now assume $m_1>0$ and $m_2<0$, that is, the point mass with mass $m_2$ is an NMO. 
When $m_1>-m_2$, that is, the orbit can be an ellipses, circle when $\frac{E_\mathrm{rel}}{\mu}<0$, the position vector of the centre of mass $\bm{R}$ lies on the straight line connecting the point mass with $m_1>0$ and the point mass with $m_2<0$ and outside of the point mass with $m_1>0$. 
In fact, if we choose $\bm{r}_1=(0,0,0)$ and $\bm{r}_2=(r,0,0)$ with $r>0$, we find $\bm{R}=\left( \frac{m_2 r}{m_1+m_2}, 0, 0\right)$. 
We should note $\frac{m_2 r}{m_1+m_2}$ is negative, $\frac{m_2 r}{m_1+m_2}<0$. 
The point mass with $m_1$ suffers a repulsive force from the point mass with $m_2$, and the force becomes a centripetal force for the point mass with $m_1$ to the centre of mass. 
Therefore, when $m_1>-m_2$ and $\frac{E_\mathrm{rel}}{\mu}<0$, two point masses with mass $m_1>0$ and $m_2<0$ make a bound state. 

We may also consider a special case $m_1=-m_2=m>0$, where the total mass $M$ in \eqref{MmuRr} vanishes and the reduced mass $\mu$ and the position vector of the centre of mass $\bm{R}$ diverge. 
Eq.~\eqref{mmtm} reduces 
\begin{align}
\label{consv}
\dot{\bm{r}}_1 - \dot{\bm{r}}_2=\bm{v}_0 \, , \quad 
\bm{r}_1 - \bm{r}_2 = \bm{v}_0 \left(t - t_0\right)\, .
\end{align}
Here $\bm{v}_0$ is a constant vector and in the second equation in \eqref{consv}, $t_0$ is a constant of the integration. 
Then the equations in \eqref{eom} have the following form 
\begin{align}
\label{eom3}
m \ddot{\bm{r}}_1 = - m \ddot{\bm{r}}_2 = - \frac{G m^2 \bm{v}_0}{\left| \bm{v}_0\right|^3 \left(t - t_0\right)^2}\, \, ,
\end{align}
whose solution is 
\begin{align}
\label{slem}
\bm{r_1} = \bm{r}_2 + \bm{v}_0 \left(t - t_0\right) = \frac{G m^2 \bm{v}_0}{\left| \bm{v}_0\right|^3} \ln \frac{t - t_0}{t_0} + \bm{v}_1 t + \bm{r}_0 \, .
\end{align}
Here $\bm{v}_1$ and $\bm{r}_0$ are vectors for the constants of the integration. 
If there are two objects before $t_0$, $t<t_0$, the second equation of Eq.~\eqref{consv} tells that the two objects collide with each other and they may vanish. 
The singular behaviour when $t\to t_0$ could indicate that Newton's approximations for gravity and mechanics could become invalid when $t\to t_0$. 
In \cite{Manfredi:2026qoz}, the above behaviours have been considered. 

\section{Can Negative Mass Objects be observed?}\label{obsrvtn}

If we explicitly write Eq.~\eqref{lmssnl} by using the speed of light $c\sim 3\times 10^8\, \mathrm{m}/\mathrm{s}$ and Newton's gravitational constant $G\sim 7\times 10^{-11}\, \mathrm{N}\mathrm{m}^2/\mathrm{kg}^2= 7\times 10^{-11}\, \mathrm{m}^3 \mathrm{s}^{-2} \mathrm{kg}^{-1}$, we find 
\begin{align}
\label{lmssnl2}
M_\mathrm{lost} 
= \frac{32\pi G^3 M^3\rho_0}{c^6} \, . 
\end{align}
If we choose $M$ as the mass of the sun, $M\sim 2\times 10^{30}\, \mathrm{kg}$. 
Then one gets $\frac{32\pi G^3 M^3\rho_0}{c^6}\sim 10^{2-30+91-3-48}\rho_0 = 10^{12} \left(\mathrm{m}^3 \right) \rho_0$. 
Then in order to satisfy the condition~\eqref{nlttlmss} that the total mass is negative, $\rho_0$ must be $\rho_0\gtrsim 10^{18}\, \mathrm{kg}\cdot\mathrm{m^{-3}}$, which is very large. 
In fact, the average density $\rho_\mathrm{univ}$ of the universe is $\rho_\mathrm{univ}\sim 10^{-26}\, \mathrm{kg}\cdot\mathrm{m^{-3}}$. 
Because by using the negative cosmological constant $\Lambda$ in \eqref{JGRG6}, we have $\rho_\mathrm{univ}\sim \rho_0 + \frac{\Lambda}{2\kappa^2}$ and therefore we need a fine-tuning of the parameters. 
In the case of the two-scalar model \eqref{I8} or the scalar-Einstein-Gauss-Bonnet gravity in \eqref{g2}, the mass of the NMOs depends on the parameters specifying the model. 

As we have seen, if there is a system composed of a positive mass object with a mass $m_1>0$ and an NMO $m_2<0$, if $m_1>-m_2$, there may be a bound state. 
Both of the objects rotate around the centre of mass. 
If there is a bright background, the light from the background could flash by the period of the rotation. 

If there is a large positive mass object, the NMOs could gather to the positive mass object, as in the nucleus and electrons in an atom. 
In such a system, the large positive mass is screened by the NMOs. 
Such a system might also be observed due to the rather strange lensing effect. 

As shown in \eqref{Trt2}, the creation of NMOs is associated with an energy flow. 
If such a flow is large, which could require that the mass $M$ should be large or the time scale $t_0$ of the creation of the NMOs should be large, we may observe such a flow. 
This flow could be distingusihed with the energy flow from black holes because NMOs have negative mass. 

Note that we have here a totally classical consideration. 
However, the possibility of the existence of bound NMOs may indicate the possibility of NMO stars or even clusters of NMOs.

\section{Object with vanishing mass}\label{Svmss}

If $\e^{2\nu}$ behaves as $\e^{2\nu}=1 + \mathcal{O}\left( r^{-2} \right)$ when $r$ is large, the ADM mass vanishes. 
We may consider the object with the vanishing mass. 
Such objects with the vanishing mass were considered and well-investigated in \cite{Chetouani:1984qdm, Perlick:2003vg, Abe:2010ap, Nakajima:2012pu, Toki:2011zu, Tsukamoto:2012xs, Tsukamoto:2013dna, Yoo:2013cia, Izumi:2013tya, Nakajima:2014nba, Bozza:2015haa, Bozza:2015wbw, Tsukamoto:2017hva, Tsukamoto:2017hva, Bozza:2017dkv, Asada:2017vxl}. 
These objects may violate the energy conditions. 

As an example, we may consider the following, 
\begin{align}
\label{vmss}
\e^{2\nu}=\e^{-2\lambda} = 1 - \frac{\gamma M^2}{r^2 + M^2}\, .
\end{align}
Here $\gamma$ and $M$ are constant
This geometry \eqref{vmss} can be realised by using a two-scalar model in Subsection~{SectionIIIA} or a scalar-Einstein-Gauss-Bonnet gravity in Subsection~{SectionIIIB}. 

In \eqref{vmss}, if $\gamma>0$, the gravitational force is attractive, although it might be weak, but if $\gamma<0$, it is repulsive. 
The geometry~\eqref{vmss} has no curvature singularity, but if $\gamma>1$, there appears a horizon at 
\begin{align}
\label{vmsshznrds}
r=r_\mathrm{h}= \sqrt{\gamma - 1}M \, ,
\end{align}
and therefore the geometry expresses a kind of black hole, although the ADM mass vanishes. 
If $\gamma<1$, there is no horizon. 

When $\gamma>0$, because the gravitational force is attractive, there might be a photon sphere. 
Now the potential $U(r)$ in \eqref{geo2g} is given by 
\begin{align}
\label{Uvmss}
U(r) = \frac{L^2}{2r^2} \left( 1 - \frac{\gamma M^2}{r^2 + M^2} \right) - \frac{E^2}{2} \, .
\end{align}
Because
\begin{align}
\label{Uvmssdsh}
U'(r) =&\, - \frac{L^2 \left\{ r^2 - \left( \gamma -1 + \sqrt{\gamma\left( \gamma - 1 \right)} \right) M^2 \right\} 
\left\{ r^2 - \left( \gamma -1 - \sqrt{\gamma\left( \gamma - 1 \right)} \right) M^2 \right\}}
{r^2 \left(r^2 + M^2\right)^2} \, .
\end{align}
Since the equation $U'(r)=0$ has a solution if $\gamma>1$, the photon sphere appears only if the object is a massless black hole, and the radius $r_\mathrm{ph}$ of the photon sphere is given by 
\begin{align}
\label{vmssps}
r = r_\mathrm{ph} = \sqrt{\gamma -1 + \sqrt{\gamma\left( \gamma - 1 \right)}} M\, .
\end{align}
We should also note that the radius vanishes in the limit of $\gamma\to 1$. 
In general, the radius of the black hole shadow $r_\mathrm{sh}$ is given by the following equation, 
\begin{align}
\label{vmssbhshw}
r=r_\mathrm{sh} = \frac{r_\mathrm{ph}}{\e^{\nu\left(r=r_\mathrm{ph}\right)}} \, .
\end{align}
Then we find 
\begin{align}
\label{vmssbhshw2}
r_\mathrm{sh} = \left( \sqrt{\gamma} + \sqrt{\gamma -1} \right) M \, .
\end{align}
An interesting point is that even if $\gamma\to 1$, where the radii of the horizon and the photon sphere vanish, $r_\mathrm{h},\ r_\mathrm{ph}\to 0$, the radius of the shadow does not vanish, $r_\mathrm{sh}\to M$. 

In the case of the Schwarzschild black hole, it is well-known that $r_\mathrm{ph}=\frac{3}{2} r_\mathrm{h}$ and $r_\mathrm{sh} = \frac{3\sqrt{3}}{2} r_\mathrm{h}$, but in the present model, we have, 
\begin{align}
\label{rdsrltn}
r_\mathrm{ph}=\sqrt{ 1 + \sqrt{\frac{\gamma}{\gamma-1}}} r_\mathrm{h}\, , \quad
r_\mathrm{sh} = \left( 1 + \sqrt{\frac{\gamma}{\gamma - 1}} \right) r_\mathrm{h}\, ,
\end{align}
which depends on the value of $\gamma>1$. 

In order to find the image of the black hole shadow, we need to know the orbit of photon. 
Instead of \eqref{geo2gB}, the equation which gives the shape of the photon orbit is given by 
\begin{align}
\label{geovmss}
0 =\frac{L^2}{2r^4} \left(\frac{dr}{d\varphi} \right)^2 + \frac{L^2}{2r^2}\left( 1 - \frac{\gamma M^2}{r^2 + M^2} \right) - \frac{E^2}{2} \, .
\end{align}
Then one gets 
\begin{align}
\label{phrvmss}
\varphi = \int \frac{L\sqrt{r^2 + M^2} dr}{E r\sqrt{r^4 + \left( M^2 - \frac{L^2}{E^2} \right) r^2 + \frac{L^2}{E^2} \left( \gamma - 1 \right) M^2}} \, . 
\end{align}
If we change the variable from $r$ to $\eta$ as follows, 
\begin{align}
\label{cvvmss}
r^2 = \frac{1}{2} \left( - M^2 + \frac{L^2}{E^2} \right) - \frac{1}{2} \sqrt{\left( M^2 + \frac{L^2}{E^2} \right)^2 - \frac{4\gamma L^2 M^2 }{E^2}} 
+ \sqrt{\left( M^2 + \frac{L^2}{E^2} \right)^2 - \frac{4\gamma L^2 M^2 }{E^2}} \cosh^2 \eta \, , 
\end{align}
we can rewrite \eqref{phrvmss}, 
\begin{align}
\label{phrvmss3}
\varphi =&\, \frac{L}{E\sqrt{\left\{ \left( M^2 + \frac{L^2}{E^2} \right)^2 - \frac{4\gamma L^2 M^2 }{E^2} \right\}
\left\{\frac{1}{2} \left( M^2 + \frac{L^2}{E^2} \right) - \frac{1}{2} \sqrt{\left( M^2 + \frac{L^2}{E^2} \right)^2 - \frac{4\gamma L^2 M^2 }{E^2}}\
\right\}
}} 
\nonumber \\
&\, \times \int \frac{d\eta}{\sqrt{1 + \frac{\sqrt{\left( M^2 + \frac{L^2}{E^2} \right)^2 - \frac{4\gamma L^2 M^2 }{E^2}}}
{\frac{1}{2} \left( M^2 + \frac{L^2}{E^2} \right) - \frac{1}{2} \sqrt{\left( M^2 + \frac{L^2}{E^2} \right)^2 - \frac{4\gamma L^2 M^2 }{E^2}}}
\cosh^2 \eta}} 
\left\{ 1 + \frac{\frac{M^2}
{\frac{1}{2} \left( - M^2 + \frac{L^2}{E^2} \right) - \frac{1}{2} \sqrt{\left( M^2 + \frac{L^2}{E^2} \right)^2 - \frac{4\gamma L^2 M^2 }{E^2}}} 
}{1 + \frac{\sqrt{\left( M^2 + \frac{L^2}{E^2} \right)^2 - \frac{4\gamma L^2 M^2 }{E^2}}}
{\frac{1}{2} \left( - M^2 + \frac{L^2}{E^2} \right) - \frac{1}{2} \sqrt{\left( M^2 + \frac{L^2}{E^2} \right)^2 - \frac{4\gamma L^2 M^2 }{E^2}}}
\cosh^2 \eta } \right\} \, .
\end{align}
If we analytically continue $\eta$ as 
\begin{align}
\label{anltccntntn}
\eta = i \left( \tau - \frac{1}{2} \right)\, ,
\end{align}
we obtain 
\begin{align}
\label{phrvmss4}
\varphi 
=&\, \frac{iL}{E\sqrt{\left\{ \left( M^2 + \frac{L^2}{E^2} \right)^2 - \frac{4\gamma L^2 M^2 }{E^2} \right\}
\left\{\frac{1}{2} \left( M^2 + \frac{L^2}{E^2} \right) - \frac{1}{2} \sqrt{\left( M^2 + \frac{L^2}{E^2} \right)^2 - \frac{4\gamma L^2 M^2 }{E^2}}\
\right\}
}} 
\nonumber \\
&\, \times \int \frac{d\tau}{\sqrt{1 + \frac{\sqrt{\left( M^2 + \frac{L^2}{E^2} \right)^2 - \frac{4\gamma L^2 M^2 }{E^2}}}
{\frac{1}{2} \left( M^2 + \frac{L^2}{E^2} \right) - \frac{1}{2} \sqrt{\left( M^2 + \frac{L^2}{E^2} \right)^2 - \frac{4\gamma L^2 M^2 }{E^2}}}
\sin^2 \tau}} 
\left\{ 1 + \frac{\frac{M^2}
{\frac{1}{2} \left( - M^2 + \frac{L^2}{E^2} \right) - \frac{1}{2} \sqrt{\left( M^2 + \frac{L^2}{E^2} \right)^2 - \frac{4\gamma L^2 M^2 }{E^2}}} 
}{1 + \frac{\sqrt{\left( M^2 + \frac{L^2}{E^2} \right)^2 - \frac{4\gamma L^2 M^2 }{E^2}}}
{\frac{1}{2} \left( - M^2 + \frac{L^2}{E^2} \right) - \frac{1}{2} \sqrt{\left( M^2 + \frac{L^2}{E^2} \right)^2 - \frac{4\gamma L^2 M^2 }{E^2}}}
\sin^2 \tau} \right\} \, .
\end{align}
The integration can be performed by using the incomplete elliptic integral of the first kind and the third kind, 
$F(\omega,k)$ and $\Pi(a; \omega,k)$, which are defined in Legendre's forms, 
\begin{align}
\label{inclement}
F(\omega,k) \equiv&\, \int^\omega_0 \frac{d\theta}{\sqrt{1 - k^2 \sin^2\theta}}\, , \nonumber \\
\Pi(a; \omega,k) \equiv&\, \int^\omega_0 \frac{d\theta}{\left( 1 - a \sin^2\theta \right) \sqrt{1 - k^2 \sin^2\theta}}\, ,
\end{align}
as follows 
\begin{align}
\label{phrvmss5}
\varphi 
=&\, \frac{iL}{E\sqrt{\left\{ \left( M^2 + \frac{L^2}{E^2} \right)^2 - \frac{4\gamma L^2 M^2 }{E^2} \right\}
\left\{\frac{1}{2} \left( M^2 + \frac{L^2}{E^2} \right) - \frac{1}{2} \sqrt{\left( M^2 + \frac{L^2}{E^2} \right)^2 - \frac{4\gamma L^2 M^2 }{E^2}}\
\right\}
}} 
\nonumber \\
&\, \times \left\{ F\left(\tau, \frac{i\sqrt{\left( M^2 + \frac{L^2}{E^2} \right)^2 - \frac{4\gamma L^2 M^2 }{E^2}}}
{\frac{1}{2} \left( M^2 + \frac{L^2}{E^2} \right) - \frac{1}{2} \sqrt{\left( M^2 + \frac{L^2}{E^2} \right)^2 - \frac{4\gamma L^2 M^2 }{E^2}}} \right) \right. \nonumber \\
&\, + \frac{iM^2}
{\frac{1}{2} \left( - M^2 + \frac{L^2}{E^2} \right) - \frac{1}{2} \sqrt{\left( M^2 + \frac{L^2}{E^2} \right)^2 - \frac{4\gamma L^2 M^2 }{E^2}}} \nonumber \\
&\, \left. 
\times \Pi \left( - \frac{\sqrt{\left( M^2 + \frac{L^2}{E^2} \right)^2 - \frac{4\gamma L^2 M^2 }{E^2}}}
{\frac{1}{2} \left( - M^2 + \frac{L^2}{E^2} \right) - \frac{1}{2} \sqrt{\left( M^2 + \frac{L^2}{E^2} \right)^2 - \frac{4\gamma L^2 M^2 }{E^2}}}; 
\tau, \frac{i\sqrt{\left( M^2 + \frac{L^2}{E^2} \right)^2 - \frac{4\gamma L^2 M^2 }{E^2}}}
{\frac{1}{2} \left( M^2 + \frac{L^2}{E^2} \right) - \frac{1}{2} \sqrt{\left( M^2 + \frac{L^2}{E^2} \right)^2 - \frac{4\gamma L^2 M^2 }{E^2}}} \right) 
\right\} \, ,
\end{align}
which gives the image of the photon sphere. 
In \eqref{phrvmss5}, the constant of the integration is absorbed into the redefinition of $\varphi$, again. 

\section{Summary and Discussion}\label{SD}

It is time to find any phenomena indicating anti-gravity. 
Instead of the construction of mathematical anti-gravity theory, which seems to be highly inconsistent currently, we speculate on effective anti-gravity, i.e. on NMOs and related phenomena. 
 
It resembles anti-gravity in some sense, but these objects could be theoretically consistent. 

In order to show that the NMOs are not so exotic, we conjecture that the NMOs naturally appear in the system of a positive point mass, cosmological fluid with negative pressure and negative cosmological constant in Section~\ref{eNMO}. 
As shown in Section~\ref{CnstrctnMG}, it is possible to construct models realising NMOs in the framework of the two-scalar model and the scalar-Einstein-Gauss-Bonnet gravity. 
Of course, this is just a formal solution of the corresponding gravitational equations. 
Nevertheless, we proposed the physical scenario where NMO is created. 
The orbit of the photon around NMO was investigated in Section~\ref{Phtn} and that of a massive particle in Section~\ref{mssv}. 
We also investigated the motion of the particles, including NMOs in Section~\ref{udrgrdtmch}. 
Especially, we showed that the bounded system composed of a positive massive object and an NMO can be created. 
Including such a bounded system, we indicated the possibility of observing NMOs in the sky. 
We also constructed a model whose solution includes the vanishing mass object. 
Its properties were clarified. 

As shown in \eqref{Trt2}, when an NMO is created, the positive energy flow to the radial direction appears, which might also be observed. 

When we consider the bound system of one positive mass object and an NMO, the positive mass object might be a black hole. 
The black holes with an accretion disk emit strong astrophysical radiation as observed in the active galactic nuclei (AGNs). 
Such a jet might be suppressed, deformed, or enhanced by suffering the repulsive force coming from the NMO. 

Even in the case of the supernova explosion, if the supernova is accompanied by NMO(s), the emitted energy flux could be strongly affected by the NMO(s). 
In any other violent events, like the merger of a black hole and a neutron star, if there are NMOs near the events, we may observe some evidence of the NMOs correcting the standard astrophysical data.

For small NWOs, they may receive the attractive force from the earth, and they may fall into the centre of the earth, but the usual matter receives the repulsive force.
Therfore the holes might be created by the falling down of the NWOs.
Such holes might be found in the rocks or any kind of crystals.

When Newton's law of gravity and Newton's laws of motion are valid, the differential cross section $\frac{d\sigma}{d\Omega}$ between two point masses with masses $m_1$ and $m_2$ is almost identical with that of the Coulomb scattering between two electric charges and given by 
\begin{align}
\label{crsssctn}
\frac{d\sigma}{d\Omega} = \left(\frac{Gm_1m_2}{4E}\right)^2 \frac{1}{\sin^4\frac{\Theta}{2}}\, .
\end{align}
Here $\Omega$ is the solid angle and $\Theta$ is the scattering angle. 
The $E$ is the total mechanical energy of the system when the origin of $E$ is chosen to be the energy of the system of two static point masses infinitely separated. 
As the differential cross section of the Coulomb scattering does not depend on the signatures of charge, the differential cross section between two point masses does not depend on the signature of the masses, that is, the cross section does not change even if one or two of the point masses are NMO(s). 
The mass of a region where the NMO exists is, however, screened. 
Therefore, if an NWO exists, although the total mass looks small, the scattering by a large mass can occur in the region. 
Final remark is that we have completely classical consideration of NMOs. 
However, careful account of quantum effects especially of quantum gravity effects may qualitatively change the above considerations.

\bibliographystyle{apsrev4-1}
\bibliography{References2}

\end{document}